\documentstyle[12pt]{article}

\advance\topmargin by -5\baselineskip 
\advance\textheight by 4\baselineskip 
\input tcilatex 
\begin{document} 
 
\thispagestyle{empty}  
\begin{titlepage} 
\begin{center} 
\vspace{0.5in} 
{\Huge Wavefunction corrections and off-forward gluon 
distributions 
in diffractive $ J/\psi $ electroproduction}\\ 
\vspace{1.0in} 
{\bf Pervez Hoodbhoy}   \\ 
Department of Physics    \\ 
Quaid-e-Azam University  \\ 
Islamabad, Pakistan. \\ 
   \end{center} 
\vspace{0.5in} 
\begin{abstract}  
Diffractive production of $J/\Psi$ particles by virtual 
photons on a proton target 
is studied with a view towards understanding two important 
corrections to the leading 
order result. First, the effect of Fermi motion of the heavy 
quarks is studied by  
performing a systematic expansion in the relative velocity, 
and a simple correction  
factor is derived. This is considerably less than estimated 
previously. Second, since  
the kinematics necessarily requires that non-zero 
momentum be transferred to the proton, off-forward gluon 
distributions are probed by 
the scattering process. To estimate the importance of the 
off-forwardness, we  
compute, in leading order perturbation theory, the extent of 
deviation from the  
usual forward gluon distribution in a quark.  
\end{abstract} 
\end{titlepage} 
\newpage 
 
\section{INTRODUCTION} 
 
There is considerable interest in diffractive 
electroproduction of $J/\Psi $ 
mesons off protons at high energies because this process is 
important for 
studying the gluon density in a proton at low values of 
Bjorken x. This 
interest stems from the simplicity of the leading order QCD 
expression for 
longitudinally polarized photons which was first derived by 
Ryskin \cite 
{Ryskin1,Ryskin2}, \label{one}  
\begin{equation} 
\frac{d\sigma }{dt}(\gamma ^{*}+P\rightarrow J/\Psi 
+P)=\frac{16\pi 
^3M\Gamma _{ee}}{3Q^6}\alpha 
_s(\stackrel{\_\_}{Q^2})[\xi g(\xi ,\stackrel{%
\_\_}{Q^2})]^2 
\end{equation} 
where, $\stackrel{\_\_}{Q^2}=\frac 14(Q^2+M^2)$, $M$ 
is the $J/\Psi $ mass, 
and $\Gamma _{ee}$ is the decay width into leptons. The 
above equation was 
derived under the assumptions of $s\gg Q^2\gg M^2\gg t$, 
and that Fermi 
motion of the quarks in the meson can be entirely 
neglected. It was further 
supposed that the gluon density appearing in eq.\ref{one} 
is that which 
would be measured in some inclusive hard process, i.e. that 
it corresponds 
to the matrix element of gluon operators between states of 
equal momentum. 
 
In this paper we shall examine the effect of relaxing two 
assumptions which 
go into eq.\ref{one}. The first is to take into account the 
correction 
arising from the Fermi motion of the $c\bar c$ pair. In the 
work of Brodsky 
et al\cite{Brodsky} this motion is contained in the vector 
meson light-cone 
wavefunction $\psi ^{\mbox{v}}(k_{\bot },x)$, a quantity 
which is in 
principle calculable from lattice QCD but whose presently 
unknown form is an 
important source of ambiguity. For example, Frankfurt et 
al\cite{Frankfurt} 
conclude that wavefunction effects suppress $J/\Psi $ 
production by a factor 
of 3 or more. However Ryskin et al\cite{Ryskin2} estimate 
a suppression 
factor of $0.4\leq F^2\leq 0.6$. The detailed shape of the 
wavefunction 
appears to be an important source of the difference. 
 
The method of treating the diffractive process, as well as 
Fermi motion 
corrections, used in this paper differs from previously used 
methods in an 
essential way. Rather than work in the infinite momentum 
frame and in the $%
A^{+}=0$ gauge, we shall choose the rest frame of the 
$J/\Psi $ and the 
Coulomb gauge for the soft gluons in the meson 
wavefunction. This is the 
natural choice for heavy-quark systems because one can 
then use systematic 
procedures, such as Non-Relativistic Quantum 
Chromodynamics\cite{Bodwin} 
(NRQCD) or the method developed in refs. 
\cite{Hafsa1,Hafsa2,Yusuf}, in 
order to evaluate quarkonium observables of interest to any 
desired level of 
accuracy. However, for the gluons in the fast moving 
proton we shall 
continue to use the $A^{+}=0$ gauge because this is the 
natural gauge to use 
for parton distributions. The two kinds of gluons have very 
different 
momenta and hence are effectively distinguishable, 
justifying the use of two 
different gauges in two different parts of the same Feynman 
diagram. It 
turns out that a gauge-invariant correction factor, derived in 
this paper, 
multiplies eq.1,  
\[ 
\left( 1+\frac 89\!\frac{\nabla ^2\phi }{M^2\phi }\right) .  
\] 
The second derivative of the wavefunction is understood to 
be evaluated at 
the origin. It is a non-perturbative quantity whose value has 
to be inferred 
from some other quarkonium processes, such as decays or 
production, 
involving large momentum transfer. In previous 
work\cite{Hafsa1,Hafsa2} its 
value was estimated,  
\[ 
\!\frac{\nabla ^2\phi }{M^2\phi }\approx -0.07  
\] 
The correction factor due to Fermi motion is therefore 
around $0.96$, a 
value considerably below the other 
estimates\cite{Ryskin2,Frankfurt}. Hence, 
the ambiguity in extracting the normalization of the gluon 
distribution may 
be under better control than anticipated so far. 
 
The second issue to be considered in this paper is the gluon 
distribution 
which appears in eq.1. Recently Ji\cite{Ji} identified 
certain twist-two 
``off-forward'' quark distributions inside the proton which, 
when measured, 
will reveal the orbital angular momentum content of the 
proton. Subsequently 
Radyushkin\cite{Radyushkin} extended the discussion to 
the off-forward or 
``asymmetric'' gluon distribution in the proton and pointed 
out that 
diffractive vector meson electroproduction necessarily 
measures this 
quantity. Here we have examined this issue further by 
considering gluon 
radiation from a quark and explicitly computed the off-
forward gluon 
distribution in a quark to leading order in $\alpha _s$. This 
enables an 
estimate to be made of the extent to which the gluon 
distribution measured 
in $J/\Psi $ diffractive production differs from that which 
would be 
measured in some inclusive process like $\gamma 
+P\rightarrow J/\Psi +X.$  
\section{FERMI\ MOTION} 
 
\subsection{Kinematics} 
 
We consider a massless proton target and the $t=0$ limit. 
Define two null 
vectors $p^\mu $ and $n^\mu $ with $p^2=n^2=0$ and 
$p\cdot n=1$,  
\begin{equation} 
p^\mu =\frac \Lambda {\sqrt{2}}(1,0,0,1),\;\;n^\mu =\frac 
1{\sqrt{2}\Lambda 
}(1,0,0,-1) 
\end{equation} 
$p^\mu $ is also the proton's momentum. Although we 
shall not need to do so 
explicitly, $\Lambda $ can be adjusted to bring the 
produced $J/\Psi $ to 
rest. The kinematic region of interest is considered to be 
$s\gg Q^2\gg M^2$%
. With the definitions,  
\begin{equation} 
\xi =-\frac{q^2}{2p\cdot q},\;\;q^2=-Q^2, 
\end{equation} 
the other momenta in Fig.1 are,  
\begin{eqnarray} 
q^\mu &=&-\xi p^\mu +\frac{Q^2}{2\xi }n^\mu  
\nonumber \\ 
P^{^{\prime \mu }} &=&p^\mu +\Delta ^\mu  \nonumber 
\\ 
\Delta ^\mu &=&-\xi p^\mu  \nonumber \\ 
K^\mu &=&\frac{\xi M^2}{Q^2}p^\mu +\frac{Q^2}{2\xi 
}n^\mu 
\end{eqnarray} 
The polarization vectors of the longitudinally polarized 
photon and $J/\Psi $ 
are, respectively,  
\begin{eqnarray} 
\varepsilon _L^\mu &=&\frac \xi Qp^\mu +\frac Q{2\xi 
}n^\mu  \nonumber \\ 
E_L &=&-\xi \frac M{Q^2}p^\mu +\frac{Q^2}{2M\xi 
}n^\mu 
\end{eqnarray} 
These obey $\varepsilon_{{L}} . \varepsilon _{{L}}=1$, 
$\varepsilon _{{L}} . 
q_{{L}}=0$, and $E_{{L}} . E_{{L}} = -1$, $E_{{L}} . 
K_{{L}} = 0$ with $%
K^2=M^2$. We have kept only leading terms and set 
$\Delta _{\bot }\approx 0$. 
 
\subsection{Diagrams} 
 
The leading order contribution to $J/\Psi $ diffractive 
production is given 
by the sum of the diagrams shown in Fig.2, to which must 
be added the 
contribution of two other diagrams that give the same 
numerical values 
because of time-reversal symmetry. Consider, by way of 
example, the first of 
these which has the expression,  
\begin{eqnarray} 
{\sl A}_1 &=&\int \frac{d^4k}{(2\pi )^4}\frac{d^4\ell 
}{(2\pi )^4}Tr[S_{\mu 
\nu }(k,\Delta )\,H_1^{\mu \nu }(k,\ell )M(\ell )],  
\nonumber  \label{amp} 
\\ 
H_1^{\mu \nu }(k,\ell ) &=&e_Q\ g^2\;\gamma ^\mu 
S_F(k+q-K/2+\ell )\gamma 
^\nu S_F(q-K/2+\ell ){\not \!\epsilon (q)}. 
\end{eqnarray} 
The perturbative part $H^{\mu \nu }(k,\ell )$ is different 
for the other 
diagram but the other factors in eq.\ref{amp} remain 
unchanged. We have not 
indicated colour explicitly in the above; its inclusion will 
amount to a 
simple factor which will be inserted at the end of the 
calculation. The 
non-perturbative information of the vector meson is 
contained in the 
Bethe-Salpeter wavefunction $M(\ell )$,  
\begin{equation} 
M(\ell )=\int d^4x\;e^{i\ell \cdot x}\langle K,E|T[\psi 
(x/2)\bar \psi 
(-x/2)]|0\rangle .  \label{BS} 
\end{equation} 
In the above, $\ell ^\mu $ and $x^\mu $ are, respectively, 
the relative 
momentum and relative distance of the $c\bar c$ pair. The 
non-perturbative 
information of the gluons in the proton is contained in 
$S^{\mu \nu }$,  
\begin{equation} 
S^{\mu \nu }(k,\Delta )=\int d^4x\;e^{i(k+\frac 12\Delta 
)\cdot x}\langle 
P^{^{\prime }}|T[A^\mu (-x/2)A^\nu (x/2)]|P\rangle 
\end{equation} 
While the diagrams in Fig.2 contain the leading order 
contribution to the 
cross-section, they also contain parts which are next to 
leading order 
(NLO). The sense in which these are to be understood as 
``higher order'' 
will be made precise later. Other diagrams will have to be 
included (see 
Fig.3) for a complete calculation at the NLO level. 
 
\subsection{Expansion} 
 
The diffractive process considered here has two large 
scales, $Q^2\gg M^2\gg 
\Lambda _{QCD}^2$. Since a $c\bar c$ system is close to 
being a 
non-relativistic coulombic bound-state, it allows for an 
expansion in powers 
of the heavy quark relative velocity. Hence it is useful to 
expand the inner 
integral in eq.\ref{amp},  
\begin{eqnarray} 
\Omega (k) &=&\int \frac{d^4\ell }{(2\pi )^4}H^{\mu \nu 
}(k,\ell )M(\ell 
)=\sum_{n=0}^\infty \Omega _n^{\mu \nu }  \nonumber \\ 
\  &=&\sum_{n=0}^\infty \frac 1{n!}\frac \partial {\partial 
\ell ^{\alpha 
_1}}\cdot \cdot \cdot \frac \partial {\partial \ell ^{\alpha 
_n}}H^{\mu \nu 
}\!\mid _{\ell =0}M^{\alpha _1\cdot \cdot \cdot \alpha _n}  
\label{exp} 
\end{eqnarray} 
where,  
\begin{eqnarray} 
M^{\alpha _1\cdot \cdot \cdot \alpha _n} &=&\int 
\frac{d^4\ell }{(2\pi )^4}%
\ell ^{\alpha _1}\cdot \cdot \cdot \ell ^{\alpha _n}M(\ell )  
\nonumber 
\label{me} \\ 
\  &=&i\partial ^{\alpha _1}\cdot \cdot \cdot i\partial 
^{\alpha _n}\langle 
K,E|T[\psi (x/2)\bar \psi (-x/2)]|0\rangle \mid _{x=0}  
\label{M} 
\end{eqnarray} 
 
The set of constants $M^{\alpha _1\cdot \cdot \cdot \alpha 
_n}$ provide a 
description equivalent to that of the original BS 
wavefunction in eq.\ref{BS}%
. The expansion eq.\ref{exp} is useful because the quarks 
are nearly on 
mass-shell: $(\frac 12K+\ell )^2\approx m^2$ implies that 
all components of $%
\ell ^\mu $ are small relative to the quark mass $m$ in the 
meson's rest 
frame and, in particular, $\ell \cdot n\sim (m/Q)^2$. In the 
large $Q^2$ 
limit this implies considerable simplification, giving a limit 
approximately 
independent of $\ell $,  
\begin{eqnarray} 
\frac 1{(k+q-K/2+\ell )^2-m^2+i\varepsilon } &\approx 
&\frac{2\xi }{Q^2}%
\frac 1{k\cdot n-\xi +i\varepsilon }, \\ 
\frac 1{(q-K/2+\ell )^2-m^2+i\varepsilon } &\approx &-
\frac 2{Q^2} 
\end{eqnarray} 
Inclusion of Fermi motion requires that we keep a 
sufficient number of 
derivatives w.r.t $\ell $ in eq.\ref{exp}. These may be 
computed using the 
simple Ward identity,  
\begin{equation} 
\frac{\partial}{\partial {\ell}_{\alpha}} S_F=-S_F 
{\gamma}^{\alpha} S_{%
{\small {F}}}, 
\end{equation} 
and the $Q^2\rightarrow \infty $ limit should be taken after 
performing the 
trace algebra. Stated in words, a differentiation of either 
propagator in eq.%
\ref{amp} w.r.t $\ell $ splits that propagator into two. Since 
we shall work 
upto $O($v$^2)$, only two derivatives of $H^{\mu \nu 
}(k,\ell )$ are needed. 
 
\subsection{Gauge Invariance} 
 
It is obvious from the occurence of the ordinary derivatives 
in eq.\ref{M}, 
or the form of the BS wavefunction eq.\ref{BS}, that gauge 
invariance has 
been violated. In earlier work on quarkonium processes\cite 
{Hafsa1,Hafsa2,Yusuf} we have encountered an identical 
situation -- the 
diagrams in Fig.2 yield expressions which are not gauge 
invariant to $O($v$%
^2)$ and one needs to consider additional diagrams, which 
are higher order 
in $\alpha _s$. These are shown in Fig.3. The gluon fields 
indicated in 
these diagrams combine with the ordinary derivatives to 
yield covariant 
derivatives, $\partial ^\alpha \rightarrow D^\alpha ,$ 
thereby restoring 
gauge invariance. In the Coulomb gauge, the contribution 
of explicit gluons 
is $O($v$^3)$ and so the reduction of the Bethe-Salpeter 
equation performed 
in ref.\cite{Keung} without explicit gluons is adequate upto 
$O($v$^2)$. We 
therefore arrive at the following gauge-invariant matrix 
elements:  
\begin{eqnarray} 
\langle K,E|\psi \bar \psi |0\rangle &=&\frac 
12M^{1/2}\!\left( \phi \!+\!%
\frac{\nabla ^2\phi }{M^2}\right)\; \!\!\!\not 
\!\!\!\:E^{*}\left( 1\!+\!%
\frac{\not \!\!\!\:K}M\right)  \nonumber \\ 
&&\!-\frac 16M^{1/2}\!\frac{\nabla ^2\phi 
}{M^2}\;{\!}\!\!\not 
\!\!\!\:E^{*}\left( 1\!-\!\frac{\not \!\!\!\:K}M\right)  
\nonumber \\ 
\langle K,E|\psi \!\!\stackrel{\leftrightarrow }{iD_\alpha 
}\bar \psi 
|0\rangle &=&\frac 13M^{3/2}\frac{\nabla ^2\phi 
}{M^2}E^{*\beta }(g_{\alpha 
\beta }+i\epsilon _{\alpha \beta \mu \nu }\gamma ^\mu 
\gamma _5\frac{K^\nu }%
M)  \nonumber \\ 
\langle K,E|\psi {}\stackrel{\leftrightarrow }{iD_\alpha 
}\stackrel{%
\leftrightarrow }{iD_\beta }\bar \psi |0\rangle &=&\frac 
16M^{5/2}\frac{%
\nabla ^2\phi }{M^2}\left( g_{\alpha \beta }-
\frac{K_\alpha K_\beta }{M^2}%
\right) {{\!}\!\!\not \!\!\!\:E^{*}\left( 1\!+\!\frac{\not 
\!\!\!\:K}%
M\right) }.  \label{me} 
\end{eqnarray} 
In the above, $\phi $ and $\nabla ^2\phi $ are the non-
relativistic 
wavefunction and its second derivative evaluated at zero 
separation. 
Inclusion of $\nabla ^2\phi $ amounts to taking the first 
step towards 
inclusion of Fermi motion. 
 
\subsection{Traces} 
 
All the ingredients are now in place for calculating the 
trace of the quark 
loops. Because we shall need only the leading twist piece, 
symmetric in $\mu  
$ and $\nu $, it will be sufficient to calculate,  
\begin{equation} 
\Omega_n= Tr \sum_{{\small {i=1,2}}} 
\Omega_n^{{\small {ii}}}=(-g_{\mu \nu 
}+p_\mu n_\nu +p_\nu n_\mu )Tr[\Omega _n^{\mu \nu }] 
\end{equation} 
for $n=0,1,2$ (n is the order of differentiation w.r.t $\ell $) 
and then 
keep only the leading order term in $O(1/Q)$. We record 
below the results of 
the calculation listing, for clarity, the relative contribution 
of only 
those diagrams which give a non-zero contribution,  
\begin{eqnarray} 
\Omega _0 &=&-\frac{4e_Q\ g^2\phi 
(0)}{M^{1/2}Q}\;2(1+2)\left( 1\!+\frac 23\!%
\frac{\nabla ^2\phi }{M^2\phi }\right) +O(1/Q^3)  
\nonumber \\ 
\Omega _1 &=&O(1/Q^3)  \nonumber \\ 
\Omega _2 &=&\frac{4e_Q\ g^2\phi 
(0)}{M^{1/2}Q}\;2(\frac 23+\frac 43+\frac 
43-\frac{8}3)\frac{\nabla ^2\phi }{M^2\phi }+O(1/Q^3)  
\label{omeg} 
\end{eqnarray} 
The factor of 2 multiplying the brackets in the above 
equations comes from 
the diagrams which are permutations of the ones shown. 
The sum over all 
diagrams is,  
\begin{equation} 
\Omega =-\frac{24e_Q\ g^2}{M^{1/2}Q}\!\phi (0)\left( 
1+\frac 49\!\frac{%
\nabla ^2\phi }{M^2\phi }\right) +O(1/Q^3).  
\label{omega} 
\end{equation} 
Note that this leading order contribution is in fact 
independent of the 
gluon momentum $k$ in the $Q^2\rightarrow \infty $ limit. 
The term in the 
brackets represents the correction due to the Fermi-motion 
of the heavy 
quarks and its square is precisely the factor which modifies 
eq.1. 
 
\section{GLUON\ DISTRIBUTION} 
 
\subsection{Asymmetric Distribution} 
 
Let us now return to the amplitude for diffractive 
scattering, a typical 
contribution to which is given by eq.\ref{amp}. The photon 
and proton both 
move along the $\hat z$ direction, and the gluons in the 
proton have limited  
$k_{\bot }^2$ and $k^2$. This means that one can perform 
a systematic 
collinear expansion in these quantities just as in the 
treatment of 
deep-inelastic scattering\cite{Ellis},  
\begin{equation} 
H^{\mu \nu }(k,\ell )=H^{\mu \nu }(k^{+},\ell )+(k-
k^{+})_\alpha \partial 
^\alpha H^{\mu \nu }(k,\ell )\mid _{k=k^{+}}+\cdot \cdot 
\cdot 
\end{equation} 
Keeping only the first, leading twist, term gives in the 
$A^{+}=0$ gauge,  
\begin{eqnarray} 
&&\ \ \int \frac{d^4k}{(2\pi )^4}S_{\mu \nu }(k,\Delta 
)\,H^{\mu \nu 
}(k,\ell )  \nonumber  \label{ccc} \\ 
\ &=&\int dy\int \frac{d\lambda \;}{2\pi }e^{i\lambda (y-
\frac 12\xi 
)}\langle P^{^{\prime }}|A_\mu (-\frac \lambda 2n)A_\nu 
(\frac \lambda 
2n)]|P\rangle \;H^{\mu \nu }(y,\ell )  \label{ccc} 
\end{eqnarray} 
In the above we have set $x^{-}=\lambda n^{-}$ and 
$k^{+}=yp^{+}.$ The time 
ordering operation becomes irrelevant on the light-cone. 
 
The inner integral will now be analyzed following the 
discussion given by 
Radyushkin\cite{Radyushkin}. Define the ``asymmetric 
distribution 
function'', $F_\xi (X),$ as below,  
\begin{eqnarray} 
&&\langle P^{^{\prime }}|n^{-}G^{+i}(-\frac \lambda 
2n)n^{-}G^{+i}(\frac 
\lambda 2n)]|P\rangle  \nonumber \\ 
\ &=&\bar u(p^{\prime }){\!}\not 
\!n\,u(p)\int\limits_0^1dX\;\left\{ 
e^{i\lambda (X-\xi /2)}+e^{-i\lambda (X-\xi /2)}\right\} 
F_\xi (X) 
\label{def} 
\end{eqnarray} 
A sum over transverse components ($i=1,2)$ is implied. 
The proton spinor 
product is $\bar u(p^{\prime }){\!}\not \!n\,u(p)=2\sqrt{1-
\xi }$, with the 
initial and final protons having the same helicity and 
$p^{\prime }=(1-\xi 
)p $. Making a Fourier transformation yields,  
\begin{equation} 
\ F_\xi (y)+F_\xi (\xi -y)=\frac 1{2\sqrt{1-\xi }}\int 
\frac{d\lambda \;}{%
2\pi }e^{i\lambda (y-\frac 12\xi )}\langle P^{^{\prime 
}}|n^{-}G^{+i}(-\frac 
\lambda 2n)n^{-}G^{+i}(\frac \lambda 2n)]|P\rangle 
\end{equation} 
It is instructive to insert a complete set of states,  
\begin{equation} 
\ F_\xi (y)+F_\xi (\xi -y)\ =-\frac{y(\xi -y)}{4\sqrt{1-\xi 
}}\sum_k\left\{ 
\delta (y-1+x)+\delta (\xi -y-1+x)\right\} \langle 
P^{^{\prime 
}}|A^i|k\rangle \langle k|A^i|P\rangle  \label{sum} 
\end{equation} 
Here $x=k\cdot n$ with $0<x<1$ is the momentum fraction 
carried by the 
intermediate state. Comparing with the usual (diagonal) 
gluon distribution 
function for $\xi =0$ it immediately follows that,  
\begin{eqnarray} 
F_{\xi =0}(y)&=&\frac 14y\,g(y),  \nonumber \\ 
g(y)&=&y\sum_k\delta (y-1+x)\langle P|A^i|k\rangle 
\langle k|A^i|P\rangle 
\label{usual} 
\end{eqnarray} 
 
We shall now relate the matrix element in eq.\ref{ccc} to 
$F_\xi (y).$ 
Inverting the relation $G^{+i}=\partial ^{+}A^i$ gives,  
\begin{equation} 
A^i(\lambda n)=n^{-}\int_0^\infty d\sigma 
\,G^{+i}(\lambda n+\sigma n). 
\end{equation} 
Inserting the above into eq.\ref{ccc} and using the 
definition of $F_\xi (y)$ 
in eq.\ref{def},  
\begin{eqnarray} 
&&\ \ \ \int \frac{d\lambda \;}{2\pi }e^{i\lambda (y-\frac 
12\xi )}\langle 
P^{^{\prime }}|\,A^i(-\frac \lambda 2n)A^i(\frac \lambda 
2n)|P\rangle  
\nonumber \\ 
\ &=&-2\sqrt{1-\xi }\left\{ \frac{F_\xi (y)}{y(\xi -y-
i\varepsilon )}+\frac{%
F_\xi (\xi -y)}{(\xi -y-i\varepsilon )(y-i\varepsilon 
)}\right\} 
\end{eqnarray} 
The imaginary part of the above for $y>0$ is  
\[ 
-2\pi \sqrt{1-\xi }\frac{F_\xi (\xi )}\xi \delta (\xi -y),  
\] 
and hence,  
\begin{equation} 
\ Im\int \frac{d^4k}{(2\pi )^4}\frac{d^4\ell }{(2\pi 
)^4}Tr[S_{\mu \nu 
}(k,\Delta )\,H^{\mu \nu }(k,\ell )M(\ell )]=-2\pi \sqrt{1-\xi 
}\frac{F_\xi 
(\xi )}\xi \Omega \ . 
\end{equation} 
 
\subsection{Cross-section} 
 
All the ingredients are now in place for calculating the 
cross-section for 
the diffractive process under consideration,  
\begin{eqnarray} 
\frac{d\sigma }{dt} &=&\frac 1{16\pi s^2}\left| A\right| ^2  
\nonumber \\ 
\ &=&\frac 1{16\pi (Q^2/\xi )^2}\left( \frac 
2{3\sqrt{3}}\right) ^2\left( 
\,2\pi \sqrt{1-\xi }\frac{F_\xi (\xi )}\xi \Omega \right) ^2 
\end{eqnarray} 
The factor $\frac 2{3\sqrt{3}}$ comes from summing over 
colours, and $\Omega  
$ is the quantity calculated in the previous section, 
eq.\ref{omega} , from 
the expansion of the heavy quark loop integral. Defining 
$\Gamma $ to be the 
leading order decay width into lepton pairs,  
\begin{equation} 
\Gamma =\frac{16\pi e_Q^2\alpha _e^2}{M^2}, 
\end{equation} 
yields the following important result,  
\begin{equation} 
\frac{d\sigma }{dt}=\frac{16\pi ^3M\Gamma 
}{3Q^6}\alpha _s(\stackrel{\_\_}{%
Q^2})\left[ 4\sqrt{1-\xi }F_\xi (\xi )\right] ^2\,\left( 1+\frac 
89\!\frac{%
\nabla ^2\phi }{M^2\phi }\right) . 
\end{equation} 
Making the approximate identification,  
\begin{equation} 
4\sqrt{1-\xi }F_\xi (\xi )\approx \xi g(\xi ),  \label{zzz} 
\end{equation} 
and setting the last factor to unity reproduces eq.1 once 
again. This 
identification was motivated by eq.\ref{usual} but the exact 
relation 
between $F_\xi (\xi )$ and $g(\xi )$ is far from clear. 
 
\subsection{Perturbative gluon distribution} 
 
$F_\xi (\xi )$ and $g(\xi )$ can be known only if the non-
perturbative 
structure of the proton state is known. However, it would 
be highly 
desirable to have at least some partial knowledge of their 
structure. To 
this end, consider the following simple solvable problem: 
imagine that the 
target proton is replaced by a single quark which can 
radiate a gluon. Its 
light-cone wavefunction can be computed order by order in 
perturbation 
theory, and the leading order matrix element is,  
\begin{equation} 
\langle P^{^{\prime }}s^{^{\prime }}|\,A^i\,|ks\rangle 
=g\frac{k^{+}}{%
p^{^{\prime }+}}\frac 1{k_{\perp }^2}\sum_\lambda \bar 
u(p^{\prime 
}s^{^{\prime }})\,\!{\!}\epsilon (l\lambda 
)\,u(ps)\,\varepsilon 
{}^{*i}(l\lambda ), 
\end{equation} 
where $l,\lambda $ are the momenta and transverse 
polarizations of the 
emitted gluon. Using,  
\begin{equation} 
\sum_\lambda \varepsilon ^\mu (l\lambda )\,\varepsilon 
^\nu {}^{*}(l\lambda 
)=-g^{\mu \nu }+\frac{l^\mu n^\nu +l^\nu n^\mu }{l\cdot 
n}, 
\end{equation} 
and summing over $i=1,2$ gives,  
\begin{equation} 
\langle P^{^{\prime }}|A^i|k\rangle \langle k|A^i|P\rangle 
=g^2\frac{2x}{%
\sqrt{1-\xi }}\frac 1{k_{\perp }^2}\frac{1+x^2-\xi }{(1-
x)(1-x-\xi )}. 
\end{equation} 
Inserting this into eq.\ref{sum} yields,  
\begin{eqnarray} 
\ \ F_\xi (y)+F_\xi (\xi -y) &=&-g^2\frac{y(\xi -y)}{2\left( 
1-\xi \right) }%
\int \frac{d^2k}{16\pi ^3k_{\perp }^2}\ \int dx  \nonumber 
\\ 
&&\ \left\{ \delta (y-1+x)+\delta (\xi -y-1+x)\right\} 
\frac{1+x^2-\xi }{%
(1-x)(1-x-\xi )}  \nonumber \\ 
\  &=&\frac{\alpha _s}{4\pi }\left\{ 1+\frac{(1-y)^2+(1-\xi 
+y)^2}{2(1-\xi )}%
\right\} \int \frac{dk_{\perp }^2}{k_{\perp }^2} 
\end{eqnarray} 
The last integral is both infrared and ultraviolet divergent. 
It is 
regulated by inserting a low momentum scale cutoff $\mu 
\sim O(\Lambda 
_{QCD})$ and a high momentum cutoff $k_{\perp }\sim 
O(Q).$ Multiplying by 
the colour factor $C_F$ $=\frac 43$, we arrive at the 
perturbative {\it %
asymmetric }gluon distribution inside a quark,  
\begin{equation} 
F_\xi (y)=\frac{\alpha _s}{6\pi }\left\{ 1+\frac{(1-
y)^2}{(1-\xi )}\right\} 
\log \frac{Q^2}{\mu ^2}  \label{FF} 
\end{equation} 
Note that,  
\begin{eqnarray} 
4\sqrt{1-\xi }F_\xi (\xi ) &=&\frac{2\alpha _s}{3\pi 
}\sqrt{1-\xi }\left\{ 
1+(1-\xi )\right\} \log \frac{Q^2}{\mu ^2}  \nonumber \\ 
\  &=&\frac{4\alpha _s}{3\pi }\left( 1-\xi +\frac 18\xi 
^2+\cdot \cdot \cdot 
\right)  
\end{eqnarray} 
but that the usual perturbative {\it symmetric} distribution, 
which can also 
be obtained by first putting $\xi =0$ in eq.\ref{FF} and 
then setting $y=\xi  
$ is,  
\begin{equation} 
\xi g(\xi )=\frac{4\alpha _s}{3\pi }\left( 1-\xi +\frac 12\xi 
^2\right) . 
\end{equation} 
Comparison of the last two formulae gives an estimate of 
the extent to which 
the asymmetric distribution departs from the symmetric one 
as $\xi $ becomes 
larger. 
 
Finally, we remark that there exists some confusion in the 
literature about 
various factors of four. First, it is claimed in the work of 
Brodsky et al%
\cite{Brodsky} that the cross-section displayed in eq.1 
must be multiplied 
by $\frac 14$. We do not find this to be the case; the result 
of Ryskin\cite 
{Ryskin1,Ryskin2} appears to be correct. A second point 
of confusion, 
unrelated to the first, is contained in the work of 
Radyushkin\cite 
{Radyushkin} who states that the usual gluon distribution, 
$g(\xi ),$ is 
approximately related to the asymmetric distribution $F_\xi 
(X)$ by the 
relation $\sqrt{1-\xi }F_\xi (\xi )\rightarrow \xi g(\xi ),$ i.e. 
without 
the factor of 4 contained in eq.\ref{usual} of this paper. 
Again, we do not 
agree with this. \newpage 
\centerline{\bf Acknowledgements} I would like to thank 
Xiangdong Ji for interesting me in the problem and 
Daniel Wyler for a discussion. This work was supported in 
part by funds 
provided by the Pakistan Science Foundation$.$

\begin{center} 
{\Large Figure Captions} 
\end{center} 
 
\begin{tabbing} 
\end{tabbing} 
{\raggedright 1. Definition of kinematic variables for 
}$J/\Psi $ 
diffractive production off a proton target by a virtual 
photon.  
\begin{tabbing} 
\end{tabbing} 
{\raggedright 2. Diagrams which give non-zero 
contribution at order }$Q^{-1}$%
{\ and }v$^0.${\ The relative weight at this order a:b is as 
1:2. Two other 
diagrams, which are numerically equal by time-reveral 
invariance, are not 
shown. The complete expression is given in 
eq.\ref{omeg}}.  
\begin{tabbing} 
\end{tabbing} 
{\raggedright 3. Diagrams which give non-zero 
contribution at order }$Q^{-1}$%
{\ and }v$^2.${\ The crosses denote connection to external 
gluons 
originating from the proton. The relative weight at this 
order a:b:c:d:e:f 
is as -1:1:-2:2:2:-4. Note that each internal gluon zero-
momentum gluon 
line, in the Coulomb gauge, is actually just a differentiation 
of the quark 
propagator. Six other diagrams, which are numerically 
equal by time-reversal 
invariance, are not shown. The complete expression is 
given in eq.\ref{omeg}}%
. 

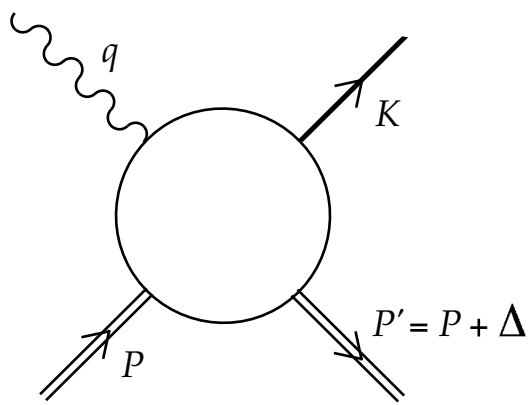
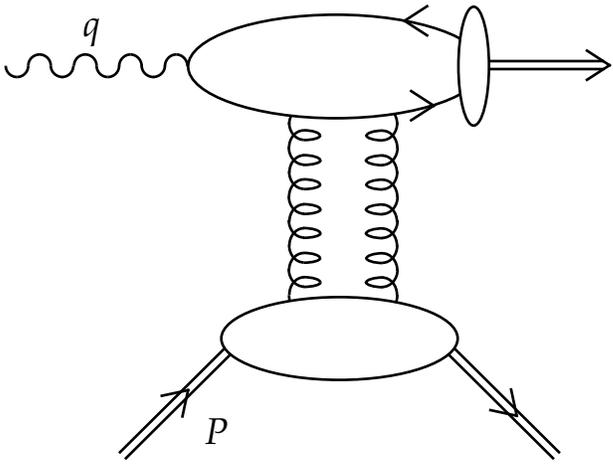
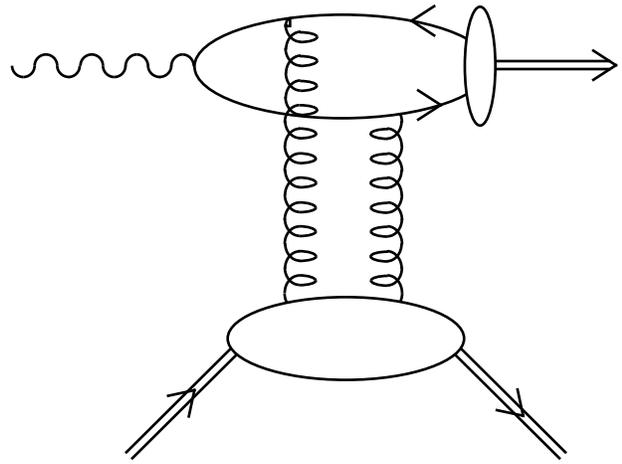
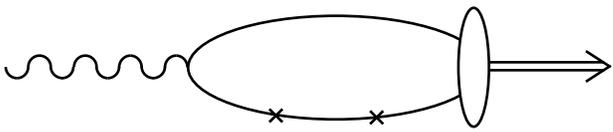
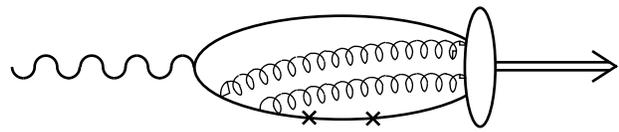
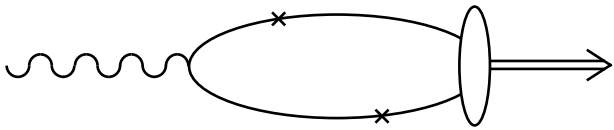
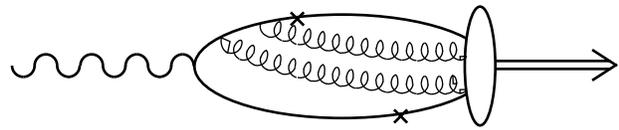
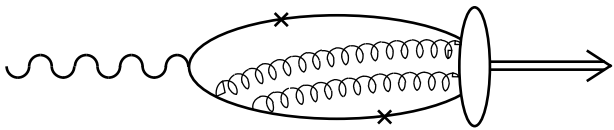
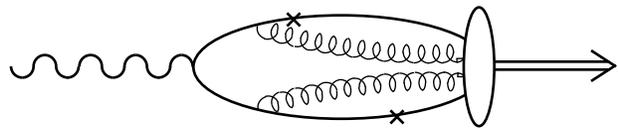
\end{document}